\documentstyle[12pt]{article}

\def\CC{\hbox{C\kern -.58em {\raise .54ex \hbox{$\scriptscriptstyle |$}}
\kern-.55em {\raise .53ex \hbox{$\scriptscriptstyle |$}} }}
\def\RR{\hbox{I\kern-.2em\hbox{R}}}
\def\sRR{{\sl \hbox{I\kern-.2em\hbox{R}}}}
\def\sqr#1#2{{\vcenter{\vbox{\hrule height.#2pt\hbox{\vrule width.#2pt
height#1pt \kern#1pt\vrule width.#2pt}\hrule height.#2pt}}}}

\def\LaTeX{{\rm L\kern-.36em\raise.3ex\hbox{\sc a}\kern-.15em
    T\kern-.1667em\lower.7ex\hbox{E}\kern-.125emX}}

\def\ie{{\it i.e.,\ }}

\def\slD{\raise.15ex\hbox{$/$}\kern-.53em\hbox{$D$}}
\def\slA{\raise.15ex\hbox{$/$}\kern-.53em\hbox{$A$}}
\def\dsl{\raise.15ex\hbox{$/$}\kern-.57em\hbox{$\Delta$}}
\def\slp{{\raise.15ex\hbox{$/$}\kern-.57em\hbox{$\partial$}}}
\def\nsl{\raise.15ex\hbox{$/$}\kern-.57em\hbox{$\nabla$}}
\def\sla{\raise.15ex\hbox{$/$}\kern-.57em\hbox{$\rightarrow$}}
\def\slla{\raise.15ex\hbox{$/$}\kern-.57em\hbox{$\lambda$}}
\def\slb{\raise.15ex\hbox{$/$}\kern-.57em\hbox{$b$}}
\def\slr{\raise.15ex\hbox{$/$}\kern-.57em\hbox{$r$}}
\def\lnp{\raise.15ex\hbox{$/$}\kern-.57em\hbox{$p$}}
\def\lnk{\raise.15ex\hbox{$/$}\kern-.57em\hbox{$k$}}
\def\lnK{\raise.15ex\hbox{$/$}\kern-.57em\hbox{$K$}}
\def\lnq{\raise.15ex\hbox{$/$}\kern-.57em\hbox{$q$}}
\def\nna{\raise.15ex\hbox{$/$}\kern-.57em\hbox{$a$}}

\def\be{{\beta}}
\def\ga{{\gamma}}
\def\de{{\delta}}
\def\eps{{\epsilon}}

\def\ze{{\zeta}}
\def\th{{\theta}}
\def\ka{{\kappa}}
\def\la{\lambda}
\def\si{{\sigma}}

\def\om{{\omega}}

\def\Th{{\Theta}}
\def\La{{\Lambda}}
\def\Si{{\Sigma}}

\def\Om{{\Omega}}

\def\bfa{{\bf a}}
\def\bfb{{\bf b}}
\def\bfc{{\bf c}}
\def\bfd{{\bf d}}
\def\bfe{{\bf e}}
\def\bff{{\bf f}}
\def\bfg{{\bf g}}
\def\bfh{{\bf h}}
\def\bfi{{\bf i}}
\def\bfj{{\bf j}}
\def\bfk{{\bf k}}
\def\bfl{{\bf l}}
\def\bfm{{\bf m}}
\def\bfn{{\bf n}}
\def\bfo{{\bf o}}
\def\bfp{{\bf p}}
\def\bfq{{\bf q}}
\def\bfr{{\bf r}}
\def\bfs{{\bf s}}
\def\bft{{\bf t}}
\def\bfu{{\bf u}}
\def\bfv{{\bf v}}
\def\bfw{{\bf w}}
\def\bfx{{\bf x}}
\def\bfy{{\bf y}}
\def\bfz{{\bf z}}
\def\bfA{{\bf A}}
\def\bfB{{\bf B}}
\def\bfC{{\bf C}}
\def\bfD{{\bf D}}
\def\bfE{{\bf E}}
\def\bfF{{\bf F}}
\def\bfG{{\bf G}}
\def\bfH{{\bf H}}
\def\bfI{{\bf I}}
\def\bfJ{{\bf J}}
\def\bfK{{\bf K}}
\def\bfL{{\bf L}}
\def\bfM{{\bf M}}
\def\bfN{{\bf N}}
\def\bfO{{\bf O}}
\def\bfP{{\bf P}}
\def\bfQ{{\bf Q}}
\def\bfR{{\bf R}}
\def\bfS{{\bf S}}
\def\bfT{{\bf T}}
\def\bfU{{\bf U}}
\def\bfV{{\bf V}}
\def\bfW{{\bf W}}
\def\bfX{{\bf X}}
\def\bfY{{\bf Y}}
\def\bfZ{{\bf Z}}

\def\pp{{\prime\prime}}
\def\gtwid{\raise.3ex\hbox{$>$\kern-.75em\lower1ex\hbox{$\sim$}}}
\def\ltwid{\raise.3ex\hbox{$<$\kern-.75em\lower1ex\hbox{$\sim$}}}
\def\12{{1\over2}}
\def\part{\partial}

\def\low#1{\lower.5ex\hbox{${}_#1$}}

\def\partt{\raise.15ex\hbox{$\widetilde$}{\kern-.37em\hbox{$\partial$}}}

\def\vev#1{{\left\langle{#1}\right\rangle}}

\def\fulltriangle{{{{{{{{{\pmb{\triangle}\kern-.65em\bullet}\kern-.4em
{\raise1.2ex\hbox{.}}}
\kern-.4em{\raise1.0ex\hbox{.}}}\kern-.2em{\raise1.0ex\hbox{.}}}
\kern-.4em{\raise.1ex\hbox{.}}}\kern-.4em{\raise.2ex\hbox{.}}}
\kern-.2em{\raise.35ex\hbox{.}}}\kern.1em{\raise.2ex\hbox{.}} }}

\def\hexagon{{\tenpoint
\langle\kern-.1em{\raise.2cm\hbox{$\overline{\hskip.7em\relax}$}}
\kern-.7em{\lower.3ex\hbox{$\underline{\hskip.7em\relax}$}}\kern-.075em
 \rangle}}

\def\pentagon{{\tenpoint
\raise.5ex\hbox{$\widehat{\qquad}$}\kern-1.8em{\backslash
\kern-.1em{\lower.3ex\hbox{$\underline{\kern.75em}$}}\kern-.05em/} }}

\def\topppageno1{\global\footline={\hfil}\global\headline
={\ifnum\pageno<\firstpageno{\hfil}\else{\hss\twelverm --\ \folio
\ --\hss}\fi}}

\def\toppageno2{\global\footline={\hfil}\global\headline
={\ifnum\pageno<\firstpageno{\hfil}\else{\rightline{\hfill\hfill
\twelverm \ \folio
\ \hss}}\fi}}

\def\boxit#1{\vbox{\hrule\hbox{\vrule\kern3pt
  \vbox{\kern3pt#1\kern3pt}\kern3pt\vrule}\hrule}}

\catcode`@=11
\def\n@me#1{\csname #1\endcsname}
\def\n@medef#1{\expandafter\edef\csname #1\endcsname}
\def\newf@m{\alloc@ 8\fam \chardef \sixt@@n} 
\def\newmathfam#1#2{
 \edef\famname{\n@me{#1fam}}
 \expandafter\newf@m\famname
 \expandafter\expandafter\expandafter\font\n@me{#1text} = #2 at 10pt
 \expandafter\expandafter\expandafter\font\n@me{#1script} = #2 at 7pt
 \expandafter\expandafter\expandafter\font\n@me{#1scriptscript} = #2 at 5pt
 \expandafter\expandafter\expandafter\textfont\n@me{#1fam} = \n@me{#1text}
 \expandafter\expandafter\expandafter\scriptfont\n@me{#1fam} = \n@me{#1script}
 \expandafter\expandafter\expandafter\scriptscriptfont\n@me{#1fam} =
\n@me{#1scriptscript}
 \n@medef{#1}{\fam=\n@me{#1fam}}
}
\catcode`\@=12
\newmathfam{acal}{eufm10}
\newmathfam{fancy}{eusm10}
\def\pmb#1{\setbox0=\hbox{$#1$}
   \kern-.025em\copy0\kern-\wd0
   \kern.05em\copy0\kern-\wd0
   \kern-.025em\raise.0433em\box0 }

\def\smallin{\hbox{{\small\rm I}\kern-.2em\hbox{{\small\rm N}}}}
\def\IN{\hbox{{\rm I}\kern-.2em\hbox{{\rm N}}}}
\def\IR{\hbox{{\rm I}\kern-.2em\hbox{{\rm R}}}}
\def\IP{\hbox{{\rm I}\kern-.2em\hbox{{\rm P}}}}
\def\IC{\hbox{{\rm I}\kern-.2em\hbox{{\rm C}}}}
\def\II{\hbox{{\rm I}\kern-.2em\hbox{{\rm I}}}}
\def\IA{\hbox{{\rm I}\kern-.2em\hbox{{\rm A}}}}
\def\ID{\hbox{{\rm I}\kern-.2em\hbox{{\rm D}}}}

\def\beginab{\begin{abstract}}
\def\endab{\end{abstract}}
\def\beginenum{\begin{enumerate}}
\def\endenum{\end{enumerate}}
\def\begindoc{\begin{document}}
\def\enddoc{\end{document}}
\def\bq{\begin{equation}}
\def\eq{\end{equation}}
\def\bqy{\begin{eqnarray}}
\def\eqy{\end{eqnarray}}
\def\bqyn{\begin{eqnarray*}}
\def\eqyn{\end{eqnarray*}}
\def\bc{\begin{center}}
\def\ec{\end{center}}
\def\bfll{\begin{flushleft}}
\def\efll{\end{flushleft}}
\def\bflr{\begin{flushright}}
\def\eflr{\end{flushright}}
\def\bigskip{\vspace{\bigskipamount}}
\def\medskip{\vspace{\medskipamount}}
\def\smallskip{\vspace{\smallskipamount}}
\def\dfr{\displaystyle\frac}
\def\dsp{\displaystyle}
\def\bskip{\baselineskip}
\def\lines{$\underline{\phantom{suzymitchellsuzymitchellsuzymitchelllll}}$}
\def\m@th{\mathsurround=0pt}
\def\n@space{\nulldelimiterspace=0pt \m@th}
\def\biggg#1{{\mbox{$\left#1\vbox to 20.5pt{}\right.\n@space$}}}
\def\Biggg#1{{\mbox{$\left#1\vbox to 23.5pt{}\right.\n@space$}}}
\def\Bigggg#1{{\mbox{$\left#1\vbox to 40pt{}\right.\n@space$}}}
\def\Biggggg#1{{\mbox{$\left#1\vbox to 50pt{}\right.\n@space$}}}
\def\Bigggggg#1{{\mbox{$\left#1\vbox to 60pt{}\right.\n@space$}}}
\def\Biggggggg#1{{\mbox{$\left#1\vbox to 80pt{}\right.\n@space$}}}
\def\noal#1{\noalign{\smallskip\noindent\rm#1\smallskip}}
\def\simsim#1{\begin{array}{c}\si\\[-7pt]\approx\end{array}}
\def\wh#1{{\widehat{#1}}}
\def\wt#1{{\widetilde{#1}}}
\def\pp#1{\frac{\p}{\p #1}}
\def\dd#1{\frac{d}{d #1}}
\def\simle{{\mathop{\stackrel{\sim}{\scriptstyle <}}\nolimits}}
\def\simgr{{\mathop{\stackrel{\sim}{\scriptstyle >}}\nolimits}}
\def\gele{{\mathop{\stackrel{>}{\scriptstyle <}}\nolimits}}
\def\lege{{\mathop{\stackrel{<}{\scriptstyle >}}\nolimits}}
\def\dotedot{\mathop{=}\limits^\cdot_{^{\scriptstyle\cdot}}}
\def\chix{{{\raise.5ex\hbox{$\chi$}}}}
\def\nms{\mathsurround=0pt}
\def\gtapprox{\mathrel{\mathpalette\overapprox>}} 
\def\ltapprox{\mathrel{\mathpalette\overapprox<}} 
\def\overapprox#1#2{\lower 2pt\vbox{\baselineskip 0pt \lineskip - 1pt
    \ialign{$\nms#1\hfil##\hfil$\crcr#2\crcr\approx\crcr}}}
\def\gtsim{\mathrel{\mathpalette\oversim>}} 
\def\ltsim{\mathrel{\mathpalette\oversim<}} 
\def\oversim#1#2{\lower 2pt\vbox{\baselineskip 0pt \lineskip 1pt
    \ialign{$\nms#1\hfil##\hfil$\crcr#2\crcr\sim\crcr}}}
\def\gtequal{\mathrel{\mathpalette\overequal>}} 
\def\ltequal{\mathrel{\mathpalette\overequal<}} 
\def\overequal#1#2{\lower 2pt\vbox{\baselineskip 0pt \lineskip 1pt
    \ialign{$\nms#1\hfil##\hfil$\crcr#2\crcr=\crcr}}}
\def\leftrightarrowfill{$\nms \mathord\leftarrow \mkern-6mu
       \cleaders\hbox{$\mkern-2mu \mathord- \mkern-2mu$}\hfill
       \mkern-6mu \mathord\rightarrow$}
\def\overdoublearrow#1{\vbox{\ialign{##\crcr
      \leftrightarrowfill \crcr\noalign{\kern-1pt\nointerlineskip}
        $\hfil\displaystyle{#1}\hfil$\crcr}}}
\def\mapright#1{\smash{\mathop{\longrightarrow}\limits^{#1}}}
\def\maprightsuper#1{\smash{\mathop{\longrightarrow}\limits^{#1}}}
\def\mapleftsuper#1{\smash{\mathop{\longleftarrow}\limits^{#1}}}
\def\maprightsub#1{\smash{\mathop{\longrightarrow}\limits_{#1}}}
\def\mapleftsub#1{\smash{\mathop{\longleftarrow}\limits_{#1}}}
\def\spmb#1{\setbox0=\hbox{$\scriptstyle#1$}
    \kern-.015em\copy0\kern-\wd0
    \kern.03em\copy0\kern-\wd0
    \kern-.015em\raise.0285em\box0 }
\def\al{\alpha}
\def\Be{\Beta}
\def\be{\beta}
\def\de{\delta}
\def\De{\Delta}
\def\ep{\epsilon}
\def\et{\eta}
\def\ga{\gamma}
\def\Ga{\Gamma}
\def\io{\iota}
\def\ka{\kappa}
\def\la{\lambda}
\def\La{\Lambda}
\def\na{\nabla}
\def\om{\omega}
\def\Om{\Omega}
\def\p{\partial}
\def\ph{\phi}
\def\Ph{\Phi}
\def\ps{\psi}
\def\Ps{\Psi}
\def\rh{\rho}
\def\si{\sigma}
\def\Si{\Sigma}
\def\ta{\tau}
\def\th{\theta}
\def\Th{\Theta}
\def\ti{\tilde}
\def\Up{\Upsilon}
\def\varep{\varepsilon}
\def\ze{\zeta}
\def\cala{{\cal A}}
\def\calb{{\cal B}}
\def\calc{{\cal C}}
\def\cald{{\cal D}}
\def\cale{{\cal E}}
\def\calf{{\cal F}}
\def\calg{{\cal G}}
\def\calh{{\cal H}}
\def\cali{{\cal I}}
\def\calj{{\cal J}}
\def\calk{{\cal K}}
\def\call{{\cal L}}
\def\calm{{\cal M}}
\def\caln{{\cal N}}
\def\calo{{\cal O}}
\def\calp{{\cal P}}
\def\calq{{\cal Q}}
\def\calr{{\cal R}}
\def\cals{{\cal S}}
\def\calt{{\cal T}}
\def\calu{{\cal U}}
\def\calv{{\cal V}}
\def\calw{{\cal W}}
\def\calx{{\cal X}}
\def\caly{{\cal Y}}
\def\calz{{\cal Z}}
\def\baroverletter#1{\setbox1=\hbox{$#1$}
  \dimen1=\ht1
    \advance\dimen1 by 1pt
  \dimen2=\wd1
    \advance\dimen2 by -1pt
  \rlap{\hspace{.5pt}\rule[\dimen1]
              {\dimen2}{.35pt}}\box1} 
\def\overA{{\baroverletter A}}
\def\overB{{\baroverletter B}}
\def\overC{{\baroverletter C}}
\def\overD{{\baroverletter D}}
\def\overE{{\baroverletter E}}
\def\overF{{\baroverletter F}}
\def\overG{{\baroverletter G}}
\def\overH{{\baroverletter H}}
\def\overI{{\baroverletter I}}
\def\overJ{{\baroverletter J}}
\def\overK{{\baroverletter K}}
\def\overL{{\baroverletter L}}
\def\overM{{\baroverletter M}}
\def\overN{{\baroverletter N}}
\def\overO{{\baroverletter O}}
\def\overP{{\baroverletter P}}
\def\overQ{{\baroverletter Q}}
\def\overR{{\baroverletter R}}
\def\overS{{\baroverletter S}}
\def\overT{{\baroverletter T}}
\def\overU{{\baroverletter U}}
\def\overV{{\baroverletter V}}
\def\overW{{\baroverletter W}}
\def\overX{{\baroverletter X}}
\def\overY{{\baroverletter Y}}
\def\overZ{{\baroverletter Z}}
\def\overxi{{\baroverletter\xi}}
\def\overka{{\baroverletter\kappa}}
\def\overal{{\baroverletter\alpha}}
\def\overbe{{\baroverletter\beta}}
\def\overGa{{\baroverletter\Gamma}}
\def\overga{{\baroverletter\gamma}}
\def\overcale{{\baroverletter{\cale}}}
\def\overvartheta{{\baroverletter\vartheta}}
\def\overth{{\baroverletter\theta}}
\def\overTh{{\baroverletter\Theta}}
\def\overvarphi{{\baroverletter\varphi}}
\def\overla{{\baroverletter\lambda}}
\def\overna{{\baroverletter\nabla}}
\def\overchix{{\baroverletter\chix}}
\def\overnu{{\baroverletter\nu}}
\def\overpsi{{\baroverletter\psi}}
\def\overPsi{{\baroverletter\Psi}}
\def\overphi{{\baroverletter\phi}}
\def\overrho{{\baroverletter\rh}}
\def\oversi{{\baroverletter\si}}
\def\overmu{{\baroverletter\mu}}
\def\overPhi{{\baroverletter\Phi}}
\def\overom{{\baroverletter\omega}}
\def\overOm{{\baroverletter\Omega}}
\def\overpar{{\baroverletter\partial}}
\def\overOm{{\baroverletter\Omega}}
\def\overDe{{\baroverletter\Delta}}
\def\overde{{\baroverletter\delta}}
\def\overtau{{\baroverletter\tau}}
\def\overze{{\baroverletter\zeta}}
\def\overep{{\baroverletter\epsilon}}
\def\overell{{\baroverletter\ell}}
\def\overeta{{\baroverletter\eta}}
\def\overa{{\baroverletter a}}
\def\overb{{\baroverletter b}}
\def\overc{{\baroverletter c}}
\def\overd{{\baroverletter d}}
\def\overe{{\baroverletter e}}
\def\overf{{\baroverletter f}}
\def\overg{{\baroverletter g}}
\def\overh{{\baroverletter h}}
\def\overi{{\baroverletter i}}
\def\overj{{\baroverletter j}}
\def\overk{{\baroverletter k}}
\def\overl{{\baroverletter i}}
\def\overm{{\baroverletter m}}
\def\overn{{\baroverletter n}}
\def\overo{{\baroverletter o}}
\def\overp{{\baroverletter p}}
\def\overq{{\baroverletter q}}
\def\overr{{\baroverletter r}}
\def\overs{{\baroverletter s}}
\def\overt{{\baroverletter t}}
\def\overu{{\baroverletter u}}
\def\overv{{\baroverletter v}}
\def\overw{{\baroverletter w}}
\def\overx{{\baroverletter x}}
\def\overy{{\baroverletter y}}
\def\overz{{\baroverletter z}}
\def\subbaroverletter#1{\setbox1=\hbox{\scriptsize$#1$}
  \dimen1=\ht1
    \advance\dimen1 by 1pt
  \dimen2=\wd1
    \advance\dimen2 by -1pt
  \rlap{\hspace{.5pt}\rule[\dimen1]
              {\dimen2}{.35pt}}\box1} 
\def\soverA{{\subbaroverletter A}}
\def\soverB{{\subbaroverletter B}}
\def\soverC{{\subbaroverletter C}}
\def\soverD{{\subbaroverletter D}}
\def\soverE{{\subbaroverletter E}}
\def\soverF{{\subbaroverletter F}}
\def\soverG{{\subbaroverletter G}}
\def\soverH{{\subbaroverletter H}}
\def\soverI{{\subbaroverletter I}}
\def\soverJ{{\subbaroverletter J}}
\def\soverK{{\subbaroverletter K}}
\def\soverL{{\subbaroverletter L}}
\def\soverM{{\subbaroverletter M}}
\def\soverN{{\subbaroverletter N}}
\def\soverO{{\subbaroverletter O}}
\def\soverP{{\subbaroverletter P}}
\def\soverQ{{\subbaroverletter Q}}
\def\soverR{{\subbaroverletter R}}
\def\soverS{{\subbaroverletter S}}
\def\soverT{{\subbaroverletter T}}
\def\soverU{{\subbaroverletter U}}
\def\soverV{{\subbaroverletter V}}
\def\soverW{{\subbaroverletter W}}
\def\soverX{{\subbaroverletter X}}
\def\soverY{{\subbaroverletter Y}}
\def\soverZ{{\subbaroverletter Z}}
\def\sovera{{\subbaroverletter a}}
\def\soverb{{\subbaroverletter b}}
\def\soverc{{\subbaroverletter c}}
\def\soverd{{\subbaroverletter d}}
\def\sovere{{\subbaroverletter e}}
\def\soverf{{\subbaroverletter f}}
\def\soverg{{\subbaroverletter g}}
\def\soverh{{\subbaroverletter h}}
\def\soveri{{\subbaroverletter i}}
\def\soverj{{\subbaroverletter j}}
\def\soverk{{\subbaroverletter k}}
\def\soverl{{\subbaroverletter l}}
\def\soverm{{\subbaroverletter m}}
\def\sovern{{\subbaroverletter n}}
\def\sovero{{\subbaroverletter o}}
\def\soverp{{\subbaroverletter p}}
\def\soverq{{\subbaroverletter q}}
\def\soverr{{\subbaroverletter r}}
\def\sovers{{\subbaroverletter s}}
\def\sovert{{\subbaroverletter t}}
\def\soveru{{\subbaroverletter u}}
\def\soverv{{\subbaroverletter v}}
\def\soverw{{\subbaroverletter w}}
\def\soverx{{\subbaroverletter x}}
\def\sovery{{\subbaroverletter y}}
\def\soverz{{\subbaroverletter z}}
\def\sovercale{{\subbaroverletter{\cale}}}
\def\soverpartial{{\subbaroverletter\partial}}
\def\soverDelta{{\subbaroverletter\Delta}}
\def\soverzeta{{\subbaroverletter\zeta}}
\def\soverth{{\subbaroverletter\theta}}
\def\sovermu{{\subbaroverletter\mu}}
\def\soverlambda{{\subbaroverletter\lambda}} 
\def\soverf{{\subbaroverletter f}} 

\def\PR{Phys.~Reports\ }
\def\APJ{Ap.J\ }
\def\JPP{J.~Plasma Phys.\ }
\def\JFM{J. Fluid Mech.\ }
\def\PL{Phys.~Lett.\ }
\def\PF{Physics of Fluids\ }
\def\PRL{Phys.~Rev.~Lett.\ }
\def\PoP{Phys.~of~Plasmas\ }
\def\PPCF{Plasma~Phys.\ }
\def\NF{Nucl.~Fusion\ }
\def\APL{Appl.~Phys.~Lett.\ }
\def\PSS{Phys.~Solid State\ }
\def\RMP{Rev. Mod. Phys.\ }

\def\support{This work was supported by the U.S.~Department of Energy
contract \mbox{No.~DE-FG05-80ET-53088} }

\def\supports{This work was supported by the U.S.~Department of Energy
contract \mbox{Nos.~DE-FG05-80ET-53088} and \mbox{DE-FG05-88ER-53266} }

\def\phys{Department of Physics\\ The University of Texas at Austin\\  Austin,
Texas~~78712}

\def\ifs{Institute for Fusion Studies\\ The
University of Texas at Austin\\ Austin, Texas~~78712}

\def\physifs{Department of Physics and Institute for Fusion Studies\\
The University of Texas at Austin\\ Austin, Texas~~78712}

\def\ifsphys{Institute for Fusion Studies and Department of Physics\\
The University of Texas at Austin\\ Austin, Texas~~78712}

\def\ifsfrc{Institute for Fusion Studies and Fusion Research Center\\
The University of Texas at Austin\\ Austin, Texas~~78712}

\def\frc{Fusion Research Center\\ The
University of Texas at Austin\\ Austin, Texas~~78712}

\def\ifspop{\small\em Institute for Fusion Studies, The
University of Texas at Austin\\ \small\em  Austin, Texas~~78712}

\def\physifspop{\small\em Department of Physics and Institute for Fusion
Studies\\
\small\em The University of Texas at Austin, Austin, Texas~~78712}

\def\ifsphyspop{\small\em Institute for Fusion Studies and Department of
Physics\\
\small\em The University of Texas at Austin, Austin, Texas~~78712}

\def\ifsfrcpop{\small\em Institute for Fusion Studies and Fusion Research
Center,
The University of Texas at Austin, Austin, Texas~~78712}

\def\frcpop{\small\em Fusion Research Center, The
University of Texas at Austin, Austin, Texas~~78712}

\def\bfA{{\bf A}}
\def\bfa{{\bf a}}
\def\bfB{{\bf B}}
\def\bfb{{\bf b}}
\def\bfC{{\bf C}}
\def\bfc{{\bf c}}
\def\bfD{{\bf D}}
\def\bfd{{\bf d}}
\def\bfE{{\bf E}}
\def\bfe{{\bf e}}
\def\bfF{{\bf F}}
\def\bff{{\bf f}}
\def\bfG{{\bf G}}
\def\bfg{{\bf g}}
\def\bfH{{\bf H}}
\def\bfh{{\bf h}}
\def\bfI{{\bf I}}
\def\bfi{{\bf i}}
\def\bfJ{{\bf J}}
\def\bfj{{\bf j}}
\def\bfK{{\bf K}}
\def\bfk{{\bf k}}
\def\bfL{{\bf L}}
\def\bfl{{\bf l}}
\def\bfM{{\bf M}}
\def\bfm{{\bf m}}
\def\bfN{{\bf N}}
\def\bfn{{\bf n}}
\def\bfO{{\bf O}}
\def\bfo{{\bf o}}
\def\bfp{{\bf p}}
\def\bfP{{\bf P}}
\def\bfq{{\bf q}}
\def\bfQ{{\bf Q}}
\def\bfR{{\bf R}}
\def\bfr{{\bf r}}
\def\bfS{{\bf S}}
\def\bfs{{\bf s}}
\def\bfT{{\bf T}}
\def\bft{{\bf t}}
\def\bfU{{\bf U}}
\def\bfu{{\bf u}}
\def\bfv{{\bf v}}
\def\bfV{{\bf V}}
\def\bfW{{\bf W}}
\def\bfw{{\bf w}}
\def\bfX{{\bf X}}
\def\bfx{{\bf x}}
\def\bfY{{\bf Y}}
\def\bfy{{\bf y}}
\def\bfZ{{\bf Z}}
\def\bfz{{\bf z}}
\newcommand{\bfchix}{{\mbox{\boldmath $\bf\chix$}}}
\newcommand{\bfell}{{\mbox{\boldmath $\bf\ell$}}}
\newcommand{\bfna}{{\mbox{\boldmath $\bf\nabla$}}}
\newcommand{\bfal}{{\mbox{\boldmath $\bf\alpha$}}}
\newcommand{\bfbe}{{\mbox{\boldmath $\bf\beta$}}}
\newcommand{\bfga}{{\mbox{\boldmath $\bf\gamma$}}}
\newcommand{\bfde}{{\mbox{\boldmath $\bf\delta$}}}
\newcommand{\bfep}{{\mbox{\boldmath $\bf\epsilon$}}}
\newcommand{\bfvarep}{{\mbox{\boldmath $\bf\varepsilon$}}}
\newcommand{\bfze}{{\mbox{\boldmath $\bf\zeta$}}}
\newcommand{\bfeta}{{\mbox{\boldmath $\bf\eta$}}}
\newcommand{\bfth}{{\mbox{\boldmath $\bf\theta$}}}
\newcommand{\bfvart}{{\mbox{\boldmath $\bf\vartheta$}}}
\newcommand{\bfio}{{\mbox{\boldmath $\bf\iota$}}}
\newcommand{\bfka}{{\mbox{\boldmath $\bf\kappa$}}}
\newcommand{\bfla}{{\mbox{\boldmath $\bf\lambda$}}}
\newcommand{\bfmu}{{\mbox{\boldmath $\bf\mu$}}}
\newcommand{\bfnu}{{\mbox{\boldmath $\bf\nu$}}}
\newcommand{\bfxi}{{\mbox{\boldmath $\bf\xi$}}}
\newcommand{\bfpi}{{\mbox{\boldmath $\bf\pi$}}}
\newcommand{\bfvarphi}{{\mbox{\boldmath $\bf\varphi$}}}
\newcommand{\bfrho}{{\mbox{\boldmath $\bf\rho$}}}
\newcommand{\bfsi}{{\mbox{\boldmath $\bf\sigma$}}}
\newcommand{\bftau}{{\mbox{\boldmath $\bf\tau$}}}
\newcommand{\bfup}{{\mbox{\boldmath $\bf\upsilon$}}}
\newcommand{\bfphi}{{\mbox{\boldmath $\bf\phi$}}}
\newcommand{\bfch}{{\mbox{\boldmath $\bf\chix$}}}
\newcommand{\bfpsi}{{\mbox{\boldmath $\bf\psi$}}}
\newcommand{\bfom}{{\mbox{\boldmath $\bf\omega$}}}
\newcommand{\bfpar}{{\mbox{\boldmath $\bf\partial$}}}
\newcommand{\bfGa}{{\mbox{\boldmath $\bf\Gamma$}}}
\newcommand{\bfDe}{{\mbox{\boldmath $\bf\Delta$}}}
\newcommand{\bfTh}{{\mbox{\boldmath $\bf\Theta$}}}
\newcommand{\bfLa}{{\mbox{\boldmath $\bf\Lambda$}}}
\newcommand{\bfXi}{{\mbox{\boldmath $\bf\Xi$}}}
\newcommand{\bfPi}{{\mbox{\boldmath$\bf\Pi$}}}
\newcommand{\bfSi}{{\mbox{\boldmath$\bf\Sigma$}}}
\newcommand{\bfUp}{{\mbox{\boldmath $\bf\Upsilon$}}}
\newcommand{\bfPhi}{{\mbox{\boldmath $\bf\Phi$}}}
\newcommand{\bfPsi}{{\mbox{\boldmath $\bf\Psi$}}}
\newcommand{\bfOm}{{\mbox{\boldmath $\bf\Omega$}}}

\newcommand{\bfcale}{{\mbox{\boldmath $\bf\cale$}}}

\newcommand{\np}{Nucl. Phys.}
\newcommand{\journal}{}

\begin{document}
\begin{flushright}
NSF-ITP-95-56
\end{flushright}
\vskip 1cm
\begin{center}
{\Large Universal relation between Green's functions in
random matrix
theory}
\vskip 1cm
{\large E. Br\'ezin}\\ {\small\em Laboratoire de Physique
Th\'eorique,
\'Ecole Normale Sup\'erieure}\\
{\small\em 24 Rue Lhomond,
75231, Paris, France}\\[.5cm]
{\small  \&}\\[.5cm]
{\large A. Zee} \\
{\small \em Institute for Theoretical Physics,}\\
{\small \em University of California, Santa Barbara, CA
93106, USA}
\vskip 1cm
\end{center}
\begin{abstract}
\bskip 24pt
We prove that in random matrix theory there exists a
universal relation between
the one-point Green's function $G$ and the connected two-
point Green's
function $G_c$ given by \vfill
$
N^2 G_c(z,w) = {\part^2 \over \part z \part w} \log (({G(z)-
G(w) \over z -w})
+ {\rm {irrelevant \ factorized \ terms.}}
$
This relation is universal in the sense that it does not
depend on the
probability distribution of the random matrices for a broad
class of
distributions, even though $G$ is known to depend on the
probability
distribution in detail. The universality discussed here
represents a different
statement than the universality we discovered a couple of
years ago, which
states that $a^2
G_c(az, aw)$ is independent of the probability
distribution, where $a$ denotes the width of the spectrum
and depends
sensitively on the probability distribution. It is shown that
the
universality proved here also holds for the more general
problem of a
Hamiltonian consisting of the sum of a deterministic term
and a random
term analyzed perturbatively by Br\'ezin, Hikami, and Zee.
\end{abstract}

\newpage
\section{Introduction}
\bskip24pt

The theory of large random matrices \cite{WIG,POR,MEH,jer}
has been extensively
developed over the years. Recently, in a series of
papers \cite{bz1,bz2,bz3,danna,bhz,hz}, we, and with the
collaboration of
J. D'Anna and of S.
Hikami, have studied the correlation between the density of
eigenvalues of large
random matrices. For physical applications, one may
imagine under some
circumstances representing the Hamiltonian of a disordered
system by a large hermitean random matrix.

We will try to make this paper as self-contained as possible
and so we will
begin with the necessary definitions. Denote by
$\varphi$ an $N$ by $N$ hermitean matrix taken from the
probability
distribution
\bq
P(\varphi)={1 \over Z}{e^{{-N}
trV(\varphi)}}\label{eq:distribution}
\eq
 with $Z$ fixed by $\int d
\varphi
P(\varphi)=1$. Here $V(\varphi)$ is an arbitrary
polynomial of $\varphi$ and the large $N$ limit is
understood. A hermitean matrix $\varphi$ is picked with
the probability
$P(\varphi)$ and we imagine that its eigenvalues are found.
When this
procedure is repeated a large number of times, we can
define a density of
eigenvalues, averaged over the matrices in an ensemble
defined by the
distribution (\ref{eq:distribution}). Similarly, the density-
density
correlation function can also be defined and calculated.

Before we continue with the necessary definitions, let us
explain the
philosophy underlying our work. We have devoted
considerable effort to
discovering and elucidating universality properties satisfied
by the
correlation function, and so it behooves us to say a few
words about our
motivations. We say a quantity is universal
if it does not depend on the probability distribution $P$, or
for the
specific class of $P$ examined here, on $V$.

While there are notable exceptions of course, much of the
literature on
random matrix theory we are aware of is concerned with
only the Gaussian
distribution, namely when $V(\varphi)$ is quadratic in
$\varphi$.
While this assumption may be plausible, it is important to
understand to
what extent the results obtained are general. Given a
Gaussian
distribution, essentially everything can be calculated. In
particular, in
the orthogonal polynomial approach, the relevant
polynomials are just the
Hermite polynomials. When asked why they restrict
themselves to the
Gaussian distribution, many workers in this subject would
simply assert
that it ``suffices." Others take the attitude that one does
what is
possible, and indeed if one were to go beyond random
matrix theory to study
actual disordered systems, it is difficult to treat non-
Gaussian disorder.

What we have been trying to do is to see, at least within the
confines of
random matrix theory, which results are indeed universal
and which results
are specific to the Gaussian case.
This search for universal quantities is made all the more
important by the
fact that the density of eigenvalues is in fact known to be
not universal
in the sense used here, that is, the density of eigenvalues
depends
sensitively on $V$, as will be mentioned below.

We would like to distinguish the universality discussed here
from the
``short distance" universality quoted in the literature. By
short distance,
we mean that we look at the correlation function on scales
larger than, but
comparable to, the spacing between the eigenvalues. On this
scale, the
physics is essentially controlled by level repulsion, leading
to a more or
less locally uniform spacing between eigenvalues. Thus, we
expect the
correlation function to be universal when suitably scaled by
the local
density (see for example equation (2.19) of \cite{bz1}). The
universality
which we have discussed in our work and which we will
discuss here is on
distance scale large compared with the spacing between the
eigenvalues and
is thus less evident.

Universality in critical phenomenon is of course well
understood by now,
essentially because of the diverging correlation length near
a second order
phase transition. The subject has been placed on a solid
foundation by the
renormalization group approach. In \cite{bz1} we suggested
that a
renormalization group inspired argument may also be
operative here,
although the details are still not completely clear,
particularly since the
density of eigenvalues is not universal. Perhaps, using the
language of
renormalization group, we may suppose that a ``relevant
operator" is
involved.

Now that we have stated some of our motivations let us
continue with our
definitions. We introduce the Green's function (or
resolvent)
\bq
G(z) = <{1\over N} tr {1\over
z - \varphi}> \equiv \int d\varphi  P(\varphi) {1\over N}tr
{1\over  z -
\varphi}
\label{eq:green} \eq
The bracket will henceforth denote averaging with respect
to $P$: for any $O(\varphi)$ we write $<O(\varphi)>=\int
d\varphi  P(\varphi) O(\varphi)$.

The
density of eigenvalues of the random matrix $\varphi$ is
then given by
$\rho(\mu)=\vev{{1\over N}tr\de(\mu-\varphi)}=-{
1\over\pi} {\rm{Im}}
G(\mu+i\eps)$.
The limit $N$ tending to infinity is always understood. Notice
that in this
paper, as in our earlier work, we choose the factors of
$N$ in our various definitions such that the interval over
which $\rho(\mu)$ is
non-zero is finite (\ie of order $N^0$) in the large $N$ limit.

The two-point Green's function is defined by
\bq
G(z,w)
\equiv\vev{{1\over N}tr{1\over z-\varphi}{1\over
N}tr{1\over
w-\varphi} }
\label{eq:1.4}
\eq
In the large $N$ limit, $G(z,w) \rightarrow G(z)G(w)$ and
thus it is
customary to define the connected Green's
function defined by $G_c(z,w) \equiv G(z,w)- G(z)G(w)$, a
quantity of order
$1/N^2$. (Thus,  we will be dealing with a quantity often
ignored
in discussions of large $N$ expansion.) The connected
correlation between
the density of eigenvalues is
then given by
\bqy
\rho_c(\mu,\nu)&=&\vev{{1\over N}tr\de(\mu-
\varphi){1\over
N}tr\de(\nu-
\varphi)}_c\nonumber\\
&=&(-1 /4\pi^2)(G_c(++)+G_c(--)-G_c(+-)-G_c(-
+)) \label{eq:1.7}
\eqy
with the obvious notation
$
G_c(\pm,\pm)\equiv G_c(\mu\pm i\eps,\nu\pm
i\de)
$
(signs uncorrelated).

Almost twenty years ago Br\'ezin, Itzykson, Parisi, and
Zuber \cite{BIPZ}
calculated the one-point
Green's function and found that, as might be expected, it
depends on $V$ in
a complicated way. Purely for the sake of completeness, let
us
record  that
for $V(\varphi)=\sum_{k=1}^p {1\over 2k}g_k \varphi^{2k}$
(we take $V$ to
be an even polynomial for  simplicity) the Green's function
has the form
\bq
G(z)={1\over 2}[V'(z)-P(z)\sqrt{z^2-a^2}]
\label{eq:form}\eq
where the polynomial
\bq
P(z)={1\over 2}\sum_{k=1}^p g_k \sum_{n=0}^{k-1}
{(2n)!\over (n!)^2}
({a^2\over 4})^n z^{2k-2n-2}
\label{eq:p}
\eq
The endpoint $a$ of the spectrum is determined by
\bq
{1\over 2}\sum_{k=1}^p g_k {(2k)!\over (k!)^2}({a^2\over
4})^k =1
\label{eq:a}
\eq
The density of eigenvalues is given by
\bq
\rho(\mu)={1\over\pi}P(\mu)\sqrt{a^2-\mu^2}
\label{eq:1.10}
\eq
In particular, in the simplest case, with the Gaussian
distribution
defined by $V(\varphi)=m^2 \varphi^2 /2$, the one-point
Green's function is
given by
\bq
G_0(z)={1\over 2}[z-\sqrt{z^2-4}]
\label{eq:gaussg0}
\eq
and thus
the density obeys
Wigner's celebrated semi-circle law
\bq
\rho(\mu)={2\over \pi a^2}{\sqrt {a^2
- \mu^2}}.
\label{eq:semicircle}\eq

We will emphatically not need any of these explicit formulas
in what
follows. We simply want to emphasize to the reader that, not
surprisingly, $G(z)$ and $\rho(\mu)$ both depend on the
potential $V$ in detail.
We say that the one-point Green's function and the density
are not universal.

In \cite{bz1} we calculated the correlation function
$\rho_c(\mu,\nu)$
using the method of
orthogonal polynomials and again as might be expected
found that it
depended in detail
on $V$. We will not display the full expression here but
simply note that
the resulting
$\rho_c(\mu,\nu)$ oscillates wildly as a function of $\mu$
and $\nu$ on
scales of $1/N$. These oscillations are entirely expected since
between
$\mu$ and $\nu$ finitely separated (that is, with $\mu-\nu
\sim O(1)$)
there are in general $O(N)$ eigenvalues.
Thus,
it is natural to smooth $\rho_c(\mu,\nu)$ by integrating
over intervals
$\delta \mu$ and $\delta \nu$,  large compared to O($N^{-
1})$
but small compared to
O($N^0$),  centered around $\mu$ and $\nu$ respectively.
Upon smoothing,
the full expression for the correlation function simplified
enormously and
we
obtained \cite{bz1}
\bq
\rho^{{\rm smooth}}_c (\mu,\,\nu)  =
{-1\over2N^2\pi^2} {1\over(\mu-\nu)^2}
{(a^2-\mu\nu)\over[(a^2-\mu^2)(a^2-\nu^2)]^{1/2}}.
\label{eq:1.16}
\eq
We found that, remarkably enough, the smoothed
correlation function depended
on the polynomial $V$ only through the single quantity $a$,
the width of
the spectrum. In other words,
if we introduce the obvious scaling variables  $x=\mu/a$
and $y=\nu/a$ then
the correlation function (henceforth we will drop the
superscript
``smooth") is equal to
\bq
\rho_c (\mu,\nu)  =
{-1\over2N^2\pi^2} {1\over a^2} f(x,y)\label{eq:uni}\eq
with the universal function
\bq
f(x,y)={1\over(x-y)^2}
{(1-xy)\over[(1-x^2)(1-y^2)]^{1/2}}
\label{eq:unifunction}
\eq
This universality has since been derived by other authors
\cite{bee,eyn,for}
using alternative methods, and verified
numerically\cite{koba}. It is
perhaps useful to remark here that,
while there are notable exceptions of course, much of the
literature on
random matrix theory, as far as we know, is devoted to the
Gaussian
case \cite{brody, verb}. The whole point of our
work is that it is possible to go beyond the Gaussian
distribution.

In \cite{bz3} we developed a diagrammatic approach to
calculating the
connected two-point Green's function $G_c(z,w)$.
We will describe this diagrammatic approach in detail in the
next section.
Here we will simply outline our result from \cite{bz3}. (To
read the rest
of this paper, it is not necessary to have read \cite{bz3}
first.)
Using a diagrammatic approach, we find that $G_c(z,w)$ is
given by an
infinite set of
Feynman diagrams. We were able to calculate and sum this
infinite set
$only$ for the Gaussian case and obtained a relatively
simple expression
for $G_c(z,w)$ (see equation (2.11) in \cite{bz3}). Later we
recognized\cite{bhz} that this expression may be written in
the elegantly
compact form
\bq
N^2 G_{0c}(z,w) = {\part^2 \over \part z \part w} \log
({G_0(z)-G_0(w)
\over z -w})
\label{eq:elegant}
\eq
The subscript ``$0$"
indicates that the quantities in this equation are all
calculated with the
Gaussian distribution. Taking the absorptive part of this
according to
(\ref{eq:1.7}) we obtain the smoothed correlation function
given in (\ref{eq:uni}). We
explained
in \cite{bz3} that the diagrammatic method ``automatically"
gives the
smoothed correlation function. This is because in the
diagrammatic method
we calculate the Green's function $G_c(z,w)$, by first letting
$N$ go to
infinity to pick out an appropriate set of diagrams, and then
letting $z$ and
$w$ approach the real axis to extract the correlation
function from the
absorptive part of $G_c(z,w)$ according to (\ref{eq:1.7}).
With $N$ going
to infinity, the discrete set
of poles of $G_c(z,w)$ on the real axis merges into a cut. This
is
equivalent to the smoothing procedure employed in
\cite{bz1}.

In \cite{bz3}, in contrast to the work we did in \cite{bz1}, we
were unable
to calculate $G_c(z,w)$ for a general $V$.
Indeed, the task of summing up the infinite sets of graphs
generated by the
interaction terms in $V$ appeared to us at the time
enormously complicated
and perhaps even hopeless. It is simple enough, however, to
summarize the
remarkable universality discovered in \cite{bz1}. The
universality
expressed in (\ref{eq:uni}) and (\ref{eq:unifunction}) can be
stated in terms
of $G_c(z,w)$
by saying that for a general $V$ we have
\bq
N^2 G_c(z,w) = {\part^2 \over \part z \part w} \log
({G_0(z/a)-G_0(w/a)
\over z -w})
\label{eq:strange}
\eq
It is far from obvious how such a relation can be derived
diagrammatically.
Here $a$ has a complicated dependence on $V$ as indicated
by (\ref{eq:a}).
Furthermore, the
Gaussian Green's function $G_0(z)$ appears on the right
hand side. The
Gaussian Green's function
$G_0(z)$,  in contrast to the Green's function
$G(z)$ appropriate to the general $V$, is not a ``natural"
object to appear
in a calculation of $G_c(z,w)$. Yet, according to the
orthogonal polynomial
analysis of \cite{bz1}, this relation must be true!

To appreciate how complicated a diagrammatic calculation of
$G_c(z,w)$ can
get, the
reader is invited to look at \cite{bhz} where together with
Hikami we
attempted this calculation. We had to restrict ourselves to
the $g
\varphi^4$ case, and even so, we were able to obtain, after a
long and
rather involved calculation, the correlation function
$G_c(z,w)$ to only
first order in
$g$. This calculation, however, was instructive. It turned out
that the
numerous terms in our final expression for the two-point
Green's function
$G_c(z,w)$ can be grouped together in precisely such a way
that the rather
complicated final expression can be written in terms of the
one-point Green's
function $G(z)$. This observation is highly non-trivial in
that, as we can see
from (\ref{eq:form}), (\ref{eq:p}), (\ref{eq:a}), the Green's
function
$G(z)$ to first order in $g$ is
already not particularly simple.

Br\'ezin, Hikami, and Zee\cite{bhz} found that, to first order
in $g$,
\bq
N^2 G_c(z,w) = {\part^2 \over \part z \part w} \log (({G(z)-
G(w) \over z
-w})(1+4gG(z)G(w))^{-1}) + O(g^2)
\label{eq:bhz}
\eq
Note that the factor of $(1+4gG(z)G(w))$ to this order in $g$
contributes to
$G_c(z,w)$ only a factorized term like $h(z)h(w)$ for some
function $h$.
These factorized terms would not contribute to the
connected correlation
function $\rho_c(\mu,\nu)$.

It was thus tempting to conjecture that for an arbitrary
$V(\varphi)$ the
connected two-point
Green's function can be written as
\bq
N^2 G_c(z,w)={\part^2 \over \part z \part w} \log (({G(z)-
G(w) \over z -w})
+ {\rm {irrelevant \  factorized \ terms}}
\label{eq:conjecture}
\eq

For the Gaussian case, the two expressions in
(\ref{eq:strange}) and
(\ref{eq:conjecture}) are manifestly the same. For a general
$V$, however,
the equality
of these two expressions is far from evident.

In a recent paper \cite{zee} we adopted a slightly different
philosophy:
instead of calculating $G(z)$ and $G_c(z,w)$ in terms of $V$
and then trying
to express $G_c(z,w)$
in terms of $G(z)$ and $G(w)$ by eliminating the
dependence on $V$, we
attempted to calculate $G_c(z,w)$ directly in terms of $G(z)$
and $G(w)$,
appealing
to $V$ only for the general structure of the Feynman
diagrams. This shift in
philosophy is reminiscent of the dispersion approach in
particle physics in
the 1950's: instead of trying to calculate various physical
quantities in
terms of an underlying Lagrangian, particle physicists of
that era
attempted to relate various physical quantities to each
other. In this
paper, we will exploit this philosophy to prove the
conjecture in
(\ref{eq:conjecture}). In line with this philosophy, we will
keep the
amount of explicit calculation to a minimum. Instead, we will
organize the
relevant diagrams in such a way as to obtain structural
relations between
different Green's functions.

\section{Diagrams}
\bskip24pt

Let us review the diagrammatic approach discussed in
\cite{bz3}.
We may regard the distribution (\ref{eq:distribution}) as
defining a $(0+0)$-dimensional field theory. In this context
the Feynman
diagram approach
consists of nothing more than expanding $G(z)$ in
inverse powers of $z$ and doing the integrals in
(\ref{eq:1.4}):
\bq
G(z) = \sum_{n=0}^{\infty} {1\over z^{n+1}} <{1\over N} tr
\varphi^n>
\label{eq:expand}
\eq
In doing the integral over $\varphi$ implied by $<.....>$ we
split off the
quadratic part $V(\varphi)$ and treat the rest of $V$
perturbatively. As
explained in \cite{bz3}, it is useful
to borrow the terminology of large $N$
quantum chromodynamics \cite{thoo}  from the
particle physics literature, and speak of quark and gluon
lines. See figure (1) for a graphical representation. (It is of
course not
necessary to use this language, and readers not familiar
with this language
can simply think of the diagrams as representing the
different terms one
encounters in doing the integral in (\ref{eq:1.4}).) The bare
quark propagator
simply comes from the explicit factors of $z$ in (\ref{eq:expand})
and is
represented
by a single line and given by $1/ z$. The quadratic term in
$V(\varphi)$ determines the bare gluon propagator,
represented by double
lines, and proportional to
\bq
\vev{\varphi^i_j \varphi^k_ l}
= \de^i_l\de^k_j{1\over Nm^2}\label{eq:2.3}
\eq
where $m^2$ is defined by the quadratic part of
$V(\varphi)=m^2
\varphi^2/2+.....$.
The non-Gaussian terms in $V(\varphi)$ describe
the interactions between gluons.

The important point, as originally stressed by 't Hooft \cite{thoo}, is
that this double-line formalism
provides an
efficient way of counting the powers of $N$ to be associated
with each
diagram: each vertex counts for one power of $N$, each
gluon propagators
counts for $N^{-1}$, and each closed loop counts for $N$.

The bare quark propagator $1/z$ is changed by the
interaction to the
dressed quark propagator $G(z)$. The gluon propagator is
dressed by gluon
interaction, but note that it is not dressed by quark loops.
This  is clear
from the definition of our problem.
Another way of saying this is to note that the one-point
Green's function
may be represented, by using the replica trick, as
\bq
G(z) = lim_{n \rightarrow 0} \int D\psi^{\dagger} D\psi
D\varphi
P(\varphi)
\psi_1^{\dagger} \psi_1
e^{-
\sum_{\alpha=1}^n \psi_{\alpha}^{\dagger} (z-
\varphi)\psi_{\alpha}}
\label{eq:rep}
\eq
The replica index $\alpha$ runs from $1$ to $n$.
Note that in this language the $\psi$'s represent the quark
fields and
$\varphi$ the gluon fields. The interaction between gluon
and quarks are
given by $ \psi_{\alpha}^{\dagger}\varphi\psi_{\alpha}$.
(Color indices are
suppressed here.) The interaction of the gluons with each
other is
determined by the non-Gaussian part of $P(\varphi)$.
Since internal quark loops are proportional to the number
of replicas $n$, they vanish in the $n\rightarrow 0$ limit.

Similarly, we can treat the two point Green's function by
expanding
\bq
G_c(z,w)=
{\sum^{\infty}_{n=1}\sum^{\infty}_{m=1}
{1\over z^{n+1} w^{m+1}}{\vev{{1\over
N}tr \varphi^n {1\over N}tr \varphi^m}}}
\label{eq:expgc}
\eq
The implied integration over $\varphi$ then generates the
Feynman diagrams for $G_c(z,w)$. We can also use the
replica trick to
represent $G_c(z,w)$. Clearly, we would have to introduce
two quark fields
$\psi_z$ and $\psi_w$: the variables $z$ and $w$ act like a
flavor label.
Thus, the correlation function describes two quarks,
``carrying" $z$ and
$w$ respectively, interacting by emitting and absorbing
gluons (which have
complicated interactions amongst themselves.) What we are
doing here may be
considered as a ``baby version" of quantum
chromodynamics.

As a warm up exercise and to gain some familiarity with
what is going on,
we will first consider
the Gaussian case. The derivation given here is simpler than
the one given
in \cite{bz3} and in
its essence was given in one of our earlier papers
\cite{lattice}. With the
benefit of hindsight, we start
by taking out two partial derivatives:
\bqy
G_c(z,w)&\equiv&\vev{{1\over N}tr{1\over z-
\varphi}{1\over
N}tr{1\over w-\varphi} }_c\nonumber\\
&=&{\part\over\part z}{\part\over\part w}{\vev{{1\over
N}tr{\log (z-
\varphi)}{{1\over N}tr{\log (w-\varphi)}}_c}}
\label{eq:logform}
\eqy

Expanding the logarithms, we find
\bq
G_c(z,w)={\part\over\part z}{\part\over\part w}
{\sum^{\infty}_{n=1}\sum^{\infty}_{k=1}
{1\over z^n w^k}{\vev{{1\over
Nn}tr \varphi^n {1\over Nk}tr \varphi^k}}}_c
\label{eq:more}
\eq
This is represented by the ``wheel" graph of Fig (2) where
the quark
propagator on the inner rim carries $z$ and the one on the
outer rim
carries $w$. For the moment we ignore quark self-energy
and  vertex
corrections: every gluon emitted on the inner rim is
absorbed on the outer
rim, and vice versa. Thus
we may set $n=k$ in (\ref{eq:more}).

Since we are working with a Gaussian distribution we can
immediately
evaluate $\vev{tr
\varphi^n tr \varphi^n}_c =
n$.  (Here with no loss of generality we have scaled $m^2$ to
unity.)
Graphically the factor of $n$ corresponds to the fact that
with the
inner rim held
fixed, we may rotate the outer rim by $n$ different ``clicks"
and leave the
diagram invariant. It is this factor of $n$ which produces the
logarithmic
function when we  evaluate the sum in (\ref{eq:more}) to
obtain
\bq
N^2G(z,w)_c=- {\part\over\part z}{\part\over\part w}\log
(1- {1\over
zw})
\label{eq:naked}
\eq
This expression does not yet have the form in
(\ref{eq:conjecture}).
Next we have to include self-energy and vertex corrections.
Instead of
doing this let us leave this expression as it is for the moment
and turn
our attention
to a ``scattering" formalism discussed in \cite{lattice}.

\section{Some formalism}
\bskip24pt

Let us go back to the
expansion of $G_c(z,w)$ in (\ref{eq:expgc}). Note that  there
is
an extra power of $1/z$ and $1/w$ compared to the powers
of $\varphi$.
Thus, in the wheel diagram of figure (2), one of the quark
propagators on
the inner rim, and one on the outer rim, should actually be
represented by
$1/z^2$ and $1/w^2$ respectively: they each consists of two
quark
propagators. We represent this fact graphically by two dots,
one on the
inner rim, and one on the outer rim, as shown in figure (3).
There are
two possibilities: the two dots are on the same ``sector" of
the wheel, as
shown in figure (3a), or the two dots are on different
``sectors", as
shown in figure (3b).

Now imagine cutting open the quark propagators at the two
dots. This
converts the wheel diagrams into two sets of scattering
diagrams as shown
in figure (4). Note that it is necessary to include the crossed
ladders in
figure (4b).  (Incidentally, the necessity of including the
crossed
ladders came to us as a bit of a surprise in carrying out this
calculation using
the formalism of \cite{bz3} but it is made completely clear
by the present
formalism.)

\section{Some formal relationships}
\bskip24pt

We have illustrated the discussion in the two preceding
paragraphs with
diagrams appropriate to the Gaussian case, but this
discussion applies
immediately to the general case. Let us define the
``scattering
amplitude"
\bqy
N\vev{({1\over{z-\varphi}})^i_j({1\over{w-
\varphi}})^m_n}_c
 &=&N(\vev{({1\over{z-\varphi}})^i_j({1\over{w-
\varphi}})^m_n}
-
\delta^i_j G(z) \delta^m_n G(w))\ \ \nonumber\\
&\equiv &\delta^i_n \delta^m_j A + \delta^i_j  \delta^m_n B
\label{eq:scattering}
\eqy

The two ``scalar" scattering amplitudes $A$ and $B$ depend
on $z$ and $w$ of
course and correspond diagrammatically  to the sets of
graphs shown in
figure (5a, b, c). We see from the flow of the color indices
that $A$ and
$B$
correspond to the two ways of cutting the wheel diagram,
that is, to figure
(4a) and figure (4b) respectively.

For a general $V$ it is complicated to calculate the
amplitudes  $A$ and
$B$ directly by perturbation theory. For example, $A$ is
given by the
infinite set of graphs in figure (5a) for the $g \varphi^4$
theory.
However, we will see that by a judicious arrangement of our
calculation, we
can avoid doing the explicit calculation that we had to work
hard to carry
out in \cite{bhz} simply to obtain a result to first order in
$g$.

First, we note that by contracting (\ref{eq:scattering}) with
$\delta^n_i
\delta^j_m$ we have
\bq
N^2 A +B
= N^2 [{{G(z)-G(w)}\over {w-z}} - G(z) G(w)]
\label{eq:abc}
\eq
We can check easily by looking at a few graphs that $A$ is of
order $N^0$
while $B$ is of order
$1/N$. Thus, we can drop $B$ in this equation and
determine $A$ (to leading
order in $N$ of course) in
terms of the one-point Green's function $G(z)$.

Next, by contracting (\ref{eq:scattering}) with $\delta^j_i
\delta^n_m$ we
find that the connected two-point Green's function is given
by
\bq
N^2 G_c(z,w)\equiv\vev{tr{1\over{z-\varphi}}tr{1\over{w-
\varphi}}}_c
=A + NB
\label{eq:twopoint}
\eq
Note that in this equation $A$ and $B$ both contributes to
the same order
in $N$. Thus, to determine
$G_c$ we still have to know $A$ and $B$. It would seem that
we would have
to work to obtain $B$, as we did in our previous papers. In
the next
section, we will see how we can
avoid calculating $B$.

\section{General $V$}
\bskip24pt

We are now ready to tackle the full problem of determining
$G_c(z,w)$ for a
general $V$. First, it is
useful to define a two-quark irreducible scattering
amplitude $\Gamma(z,w)$
consisting of those graphs that
do not fall into two disconnected pieces upon cutting the two
separate
quark propagators,
as shown in figure (6). Then the scattering function $A$ is
evidently given by
\bq
A ={1\over z^2 w^2} (\Gamma + \Gamma{1\over
zw}\Gamma + \Gamma{1\over
zw}\Gamma {1\over zw}\Gamma + .....)
={  {1\over z^2 w^2} \Gamma \over 1 - {1\over
zw}\Gamma}
\label{eq:aandgamma}
\eq
Note that this expression merely relates $A$ to $\Gamma(z,w)$, which at
this stage would appear to be an exceedingly complicated object to
calculate directly.

Let us now start the computation of $G_c(z,w)$ with (\ref{eq:expgc}) which
we repeat here for
convenience:
\bq
G_c(z,w)=
{\sum^{\infty}_{n=1}\sum^{\infty}_{m=1}
{1\over z^{n+1} w^{m+1}}{\vev{{1\over
N}tr \varphi^n {1\over N}tr \varphi^m}}}
\label{eq:expgc1}
\eq
Again, let us proceed by first ignoring vertex and self
energy corrections.
But in the general case we  can no longer simply set $n$
equal to $m$, as
we did in the Gaussian case. We see, however, that we can
express all the
wheel diagrams representing (\ref{eq:expgc}) in terms of
the (unknown) amplitude
$\Gamma(z,w)$ as indicated in figure (7).

We have yet to put on the two dots, one on the inner rim,
one on the outer
rim, as explained above. We see that now there are a
number of
possibilities. We can put the dot on an ``exposed" quark line,
or we can
put the dot on a quark line hidden inside a $\Gamma$. These
two
possibilities are illustrated in figure (8).

 For obvious reasons, we now find it useful to use
an alternative notation in which we replace $z$ and $w$  by
$z_1$ and
$z_2$ respectively and define $\part_a \equiv {\part \over
\part z_a}$.
When we put the dot on an exposed quark line, we replace
$1/z_a$ by
$1/{z_a}^2 = -\part_a (1/z_a)$.
When we put the dot on a ``hidden" quark line, we in effect
replaced
$\Gamma(z_1, z_2)$ by $-\part_a \Gamma(z_1, z_2)$.
Incidentally, in the
Gaussian case there is no ``hidden" quark line: all quark
lines are exposed
by definition. The differential operator $\part_1 \part_2$
associated with
putting on the dots is precisely what relates (\ref{eq:expgc})
to
(\ref{eq:more}).

In addition to the choice of putting the two dots on exposed
or hidden
quark lines, we also have the choice of putting the two dots
in the same
sector or
in different sectors. (Sectors are defined as the segments of
the wheel
divided by the different $\Gamma$'s: each sector consists of one the spaces
between the $\Gamma$'s and a $\Gamma$ next to that space.)

Let us first consider putting the two dots in the same sector.
Then we have
the following four possibilities, corresponding to the four
diagrams in
figure (9): (a) both dots are on an exposed line, thus giving
${\Gamma\over z_1^2 z_2^2}$, (b) the dot on the inner rim
is on a hidden
line, while the dot on the outer rim is on an exposed line,
thus giving
${-\part_1\Gamma\over z_1 z_2^2}$, (c) the previous case
with inner and
outer exchanged, and (d) both dots are on a hidden line,
thus giving
${\part_1\part_2\Gamma\over z_1 z_2}$. These four terms add up to
$({\Gamma\over z_1^2 z_2^2}
+
{-\part_1\Gamma\over z_1
z_2^2}+
{-\part_2\Gamma\over z_1^2 z_2}
+
{\part_1\part_2\Gamma\over z_1
z_2}
)
= \part_1 \part_2 ( {\Gamma \over z_1 z_2}  )$.

The rest of the
wheel (see figure
(9)) can be filled with nothing, one $\Gamma$, two
$\Gamma$'s, and so on,
that is, with the series $1+{\Gamma\over z_1 z_2}
+({\Gamma\over z_1
z_2})^2+....$. Putting all of this together we have the following
contribution to $G_c(z_1,
z_2)$:
\bq
({1\over 1-{\Gamma\over z_1 z_2}})
\part_1 \part_2 ( {\Gamma \over z_1 z_2}  )
\label{eq:samesector}
\eq

Next we have to consider the possibilities of putting the two
dots on two
different sectors. The dot on the inner rim can be either on
an exposed
line or a hidden line, and thus we obtain a factor
$({\Gamma\over z_1^2
z_2}-{\part_1 \Gamma\over z_1 z_2}) = -\part_1({\Gamma\over z_1 z_2})$.
Similarly, the dot on
the outer rim
can be either on an exposed line or a hidden line, and we
obtain the factor
just given but with $1$ and $2$ interchanged. These two factors combine to
give $\part_1({\Gamma\over z_1 z_2})\part_2({\Gamma\over z_1 z_2})$

The two
different sectors,
where the two dots are placed, divide the wheel into two
segments, each of
which can be filled, just as above, with nothing, one $\Gamma$, two
$\Gamma$'s, and so on,
that is, each segment leads to the factor ${1\over 1-
{\Gamma\over z_1
z_2}}$. Thus, we
obtain in $G_c(z_1, z_2)$ the contribution
\bq
({1\over 1-{\Gamma\over z_1 z_2}})^2
\part_1({\Gamma\over z_1 z_2})\part_2({\Gamma\over z_1 z_2})
\label{eq:differentsectors}
\eq

Putting (\ref{eq:samesector}) and (\ref{eq:differentsectors})
together, we
finally obtain
\bqy
N^2 G_c(z_1, z_2)&=&
({1\over 1-{\Gamma\over z_1 z_2}})
\part_1 \part_2 ( {\Gamma \over z_1 z_2}  )\nonumber\\
&+&
({1\over 1-{\Gamma\over z_1 z_2}})^2
\part_1({\Gamma\over z_1 z_2})\part_2({\Gamma\over z_1 z_2})
\label{eq:together}
\eqy
which we happily recognize as just
\bq
N^2 G_c(z_1, z_2)=
-\part_1\part_2 \log(1-{\Gamma\over z_1 z_2})
\label{eq:baregc}
\eq

Note that we could have given a shorter derivation by
simply ``working
backwards": we could have started with (\ref{eq:baregc})
and simply said that
the operator $(\part_1\part_2 \log)$ distributes the two
dots in all the
possible ways that we had enumerated, but we believe that
our longer derivation
just given is more transparent and easier for the reader to
follow.

For $\Gamma=1$ we recover our previous expression
(\ref{eq:naked}): surely
we are on the right track.
We are almost there but we have yet to put in the quark
self energy
corrections and the vertex corrections. The quark self
energy corrections
are easy to put in: we simply replace the bare quark
propagator $1/z$ by
the dressed propagator $G(z)$ appropriate for the
interaction potential
$V$. The vertex corrections require more thought.
First, note that our usage of the term ``vertex corrections"
differs
slightly from the standard usage. For example, the diagram
in figure (10a)
has already been counted in $\Gamma$. We include in
vertex corrections the
diagrams in figure (10b) for example. As seen in figure
(10c), including
the vertex corrections we simply multiply the amplitude
without vertex
corrections by a factor $v(z)$.

Already, we noted in \cite{bz3} that the gluons in the
vertex corrections
must ``span" the whole amplitude \cite{doug} lest we lose
factors of $N$.
This is illustrated in figure (11). The remark given in
\cite{bz3}
concerning the vertex corrections for the
Gaussian case clearly generalizes to the case of an arbitrary
$V$.

We are now faced with the task of calculating $v(z)$, which
we calculated
for the
Gaussian case in \cite{bz3}.
Fortunately, we can avoid doing any work by noting that
there is a ``Ward
identity"
\bq
v(z)={dG^{-1}\over dz}
\label{eq:ward}
\eq
The reader can easily convince himself or herself of this
identity by
contemplating the diagrams in figure (12). The
differentiation $d\over dz$
simply puts the vertex in all possible places with the correct
counting
factor. Note that the $v(z)$ given by this identity is fully
dressed, that
is, the quark propagators that enter in $v(z)$ are already
dressed.

But now we see a remarkable cancellation of the vertex
corrections because
we have the foresight (or hindsight!) of arranging our
calculation of $G_c$
so that it has the form $\part_1\part_2(....)$
before self energy and vertex corrections are included (see
(\ref{eq:baregc})). Including these
corrections we obtain
\bqy
N^2 G_c(z_1, z_2)& =& v(z_1)v(z_2) ([\part_1\part_2 \log(1-
{\Gamma \over z_1
z_2})]|_{dressed} ) \nonumber\\
&=& \part_1\part_2 (\log(1- {\Gamma \over z_1
z_2}|_{dressed}))
\label{eq:cancel}\eqy
since $({d\over dz})|_{dressed}={d\over dG^{-1}(z)}={1\over
v(z)}{d\over
dz}$. The vertex corrections disappear!

At this point, if we go back to the Gaussian formula
(\ref{eq:naked}) we
see that  in the Gaussian case we have essentially finished
our calculation
since $\Gamma=1$. We obtain
\bq
N^2G_{0c}(z,w)= - {\part\over\part z}{\part\over\part
w}\log (1-
G_0(z)G_0(w))
\label{eq:dressed}
\eq
Inserting the explicit form for $G_0(z)$ given in
(\ref{eq:gaussg0})
we obtain after some simple manipulations
(\ref{eq:elegant}).
This derivation is simpler than that given in (\cite{bz3}).

It is always nice to recover the Gaussian result as a check
but here we
want to do the much more ambitious problem of calculating
$G_c$ for a
general $V$. To go further, we have to calculate
$\Gamma(z_1, z_2)$ and
then to dress it by
replacing $1/z_a$ by $G(z_a)$ which we will write as $G_a$
for short. This
would have been a long involved calculation, but again we
note happily that
we can avoid doing it
simply by noting that the bare $\Gamma$ is related to the
bare $A$ by
(\ref{eq:aandgamma}):
\bq
\Gamma= {z_1 z_2 A \over A +{1\over z_1 z_2} }
\label{eq:baregamma}
\eq
But the dressed $A$ is given by the identity in
(\ref{eq:abc})! Thus, we don't have to do any further work. We simply dress
(\ref{eq:baregamma}) to obtain
\bq
\Gamma_{dressed} = ({1\over G_1 G_2}+{G_1 -G_2 \over z_1
-z_2})
\label{eq:dressgamma}
\eq
Finally, then
\bq
(z_1 z_2 -\Gamma(z_1, z_2))_{dressed} = ({1\over G_1
G_2}-\Gamma_{dressed})= -({z_1 -z_2 \over G_1 -G_2})
\label{eq:tttt}\eq
And thus, we have proved our conjecture. Combining
(\ref{eq:cancel}) and
(\ref{eq:tttt}) we obtain our conjectured relation
\bq
N^2 G_c(z,w)= {\part^2 \over \part z \part w} \log (({G(z)-
G(w) \over z -w})
 + {\rm {irrelevant \ factorized \
terms}}\label{eq:conjecturetrue}
\eq

Finally, in the language of the wheel diagram, it is easy to
see where the
irrelevant factorized terms in (\ref{eq:conjecturetrue})
come from. They
come from diagrams which disconnect the inner rim and
the outer rim of the
wheel from each other and thus clearly has a factorized
dependence on $z$
and $w$. See figure (13).

\section{Deterministic plus random}
\bskip24pt

In our earlier work, we have also generalized the problem
outlined in the
introduction of this paper to the problem of a Hamiltonian
given by
the sum of a deterministic term and a random term
\bq
H= H_0 + \varphi
\label{eq:det}
\eq
Here $H_0$ is a diagonal matrix
with diagonal
elements $\epsilon_i$, $i=1,2,...N$, and $\varphi$ a random
matrix taken from the ensemble (\ref{eq:distribution}). For
the
Gaussian case, namely with
$V(\varphi)={1\over2}\varphi^2$, Pastur \cite{PAS} has
long ago determined the density of eigenvalues. Our work in
\cite{bz3} went beyond Pastur's work in that the
correlation function between the density of eigenvalues in
the Gaussian
case was also determined. In our recent work with
Hikami, \cite{bhz} we studied this
correlation function for a $g\varphi^4$ theory to first order
in $g$.

This problem of ``determinism plus chance" may be
regarded
as a generic problem in physics, and as such represents a
significant generalization of Wigner's problem. For
example, consider an electron moving in a magnetic field
and scattering off
impurities. We note that these ``deterministic plus random"
problems
are considerably more difficult than purely random
problems. The orthogonal
polynomial approach used in \cite{bz1}  involves
diagonalizing
the random matrix $\varphi$ and is clearly no longer
available:
in (\ref{eq:det}) we cannot diagonalize $\varphi$ without
un-diagonalizing $H_0$. Thus, we do not have the analog of
(\ref{eq:strange}) for
this problem.

Indeed, as discussed in a recent paper\cite{zee}, the
problem described
here represents
a special case of a broader class of problems involving the
addition of random matrices. The deterministic Hamiltonian
$H_0$ may in turn be replaced by a random Hamiltonian.
Indeed, a deterministic matrix is but a special case of a
random matrix.

Consider then a
Hamiltonian given by
\bq
H=\varphi_1 +\varphi_2
\label{eq:ham}
\eq
with the matrices $\varphi_{1,2}$ taken from a factorized
probability distribution
\bq
P(\varphi_1, \varphi_2)={1 \over Z}{e^{{-N}
tr [V_1(\varphi_1) + V_2(\varphi_2)]}} \equiv
P_1(\varphi_1)P_2(\varphi_2).
\label{eq:distributionproduct}
\eq
In \cite{zee} it was shown how the one-point Green's
function $G$ can be
obtained for the Hamiltonian given in (\ref{eq:ham}). Here
we would like to
solve the problem of determing the connected two-point
Green's function
$G_c$.

A slightly sloppy but essentially correct argument is that
given our
universal relation (\ref{eq:conjecturetrue}) between $G_c$
and $G$ our
problem is solved instantly.
The desired connected two-point Green's function is given
in terms of the
one-point Green's
function appropriate to the distribution in
(\ref{eq:distributionproduct}).
This result is precisely what was conjectured in \cite{bhz}.

We can put this argument on a more solid footing by using
the formalism
discussed in \cite{zee}. (The following discussion will be
sketchy and not
self contained.) In that work, it was shown that $G(z)$ may
be
determined in terms of $G_1(z)$ and $G_2(z)$, the Green's
functions
corresponding to the distribution $P_1$ and $P_2$
respectively, according
to the following procedure. First, solve the equations
$G_a(B_a(z))=z$, for
$a=1,2$, that is, find the functional inverses of $G_a(z)$,
denoted by
$B_a(z)$ here. Next, define the function
\bq
B_{1+2}(z)=B_1(z)+B_2(z)-{1\over
z}.
\label{eq:addition}\eq
The functional inverse of $B_{1+2}(z)$ is then the desired
Green's
function $G(z)$. This type of addition laws has been
discussed recently in
the mathematical \cite{voi,hag,waterloo} and physical
literature \cite{douglas,gross,li}.

Following the argument given in \cite{zee} which we won't
repeat here, we
find for example  the undressed $\Gamma$ is given by
\bq
\Gamma(z,w)=zw(1- {z^2 G_{gc1}(z)-w^2 G_{gc1}(w) \over
z-w})+(1
\leftrightarrow  2)
\label{eq:addforgamma}\eq
Here $G_{gc1}$ and $G_{gc2}$ are ``gluon connected"
Green's functions defined in \cite{zee} and are related to
$B_1$ and $B_2$
respectively. Following the same steps as above we find that
(\ref{eq:cancel}) still holds. Inserting the expression for
$\Gamma$ given
here and using (\ref{eq:addition}) we immediately find that
(\ref{eq:conjecturetrue}) indeed holds for this more general
class of
problems.

\section{Conclusion}
\bskip24pt

In conclusion, we have found a remarkable universal
relation between the
one point Green's function $G(z)$ and the connected two
point Green's
function $G_c(z,w)$. This represents an entirely different
sort of
universality as the one found in \cite{bz1}. There it was
shown that the
scaled two point Green's function $a^2 G_c(az, aw)$,  with
$a$ the endpoint
of the spectrum given by a complicated function of the
potential
$V(\varphi)$, is independent of $V$. Here it is shown that
the structural
relation between $G_c(z,w)$ and $G(z)$ is independent of
$V$, even though
$G(z)$ is known to depend on $V$ in a complicated way.

While we know that these two forms of universality must be
equivalent, it
is not obvious how to show this equivalence directly.

The compact form of $G_c(z,w)$ obtained here renders the
universality
property of $\rho^{smooth}_c(\mu, \nu)$ as $\mu$
approaches $\nu$
particularly transparent. Consider our universal form
\bq
N^2 G_c(z,w)={\part\over \part z}{\part\over \part w} \log
({G(z)-G(w)
\over z-w})\label{eq:univform}\eq
If $z$ and $w$ approach each other on
the same side of the cut of $G$, the argument of the
logarithm is a smooth
function of $z-w$, and thus would not contribute
to $\rho^{smooth}_c(\mu, \nu)$ a term proportional to $
1/(\mu-\nu)^2$ that
we know
from (\ref{eq:1.16}) must be there. On the other hand, if $z$
and $w$
approach each other from opposite sides of the cut, then
writing $G(\mu \pm
i\epsilon) \equiv R(\mu) \pm i I(\mu)$ as $\epsilon$ goes to
zero, we have
the universal singular behavior
\bqy
N^2 G_c(\mu+i\epsilon, \nu-i \delta)&=&{\part\over \part
\mu}{\part\over
\part \nu} \log ({R(\mu)-R(\nu) \over \mu-\nu} + i
{I(\mu)+I(\nu) \over
\mu-\nu})\nonumber\\
&\rightarrow&
-{\part\over \part \mu}{\part\over \part \nu}\log (\mu-
\nu)
=-{1\over (\mu-\nu)^2 }
\label{eq:sing}\eqy
Inserting this into (\ref{eq:1.7}) we find the singular part of
(\ref{eq:1.16}). The emergence of the universal behavior, in
which the
dependence on
$R(\mu)$ and $I(\mu)$ drops away, is made particularly
clear by
(\ref{eq:sing}).

We may be tempted to conjecture that the elegantly
compact form of our
universal relation (\ref{eq:univform})
may be associated with a deeper mathematical
structure.

\section{Acknowledgement}

EB and AZ would like to thank respectively the Institute for
Theoretical Physics, Santa Barbara, and the \'Ecole Normale
Sup\'erieure,
Paris, where part of this work was done, for their
hospitality.  This work
was supported in part by the
National Science
Foundation under Grant No. PHY89-04035 and by the
Institut
Universitaire de France.

\section{Figure Captions}

Fig 1. Diagrammatic rules: (a) quark propagator, (b) gluon
propagator, (c)
quark gluon vertex, and (d) gluon interaction, illustrated
here with a
$g\varphi^4$ vertex.

Fig 2. Wheel diagram.

Fig 3. The two dots may be placed in the same sector (a) or
in different
sectors (b).

Fig 4. The diagrams in Fig 3.  cut open at the dots. (The
diagrams in the
same topological class, rather than the exact
correspondents, are shown.)

Fig 5. Diagrams contributing to $A$ (a) and diagrams
contributing to $B$
(b) and (c). Note that two topological distinct classes of
graphs
contribute to $B$.

Fig 6. Typical diagrams contributing to $\Gamma$.

Fig 7. The wheel diagrams for a general potential $V$.

Fig 8. A dot may be put on an exposed line (a) or on a
hidden line (b).

Fig 9. (a) The two dots are both placed on  exposed lines. (b)
The dot on
the outer rim is placed on an exposed line, while the dot on
the inner rim
is placed on a hidden line. (c) The situation in (b) reversed.
(d) The two
dots are both placed on hidden lines.

Fig 10. (a) Diagram not included in what we called vertex
correction. (b)
Diagram included in what we called vertex correction. (c) A
vertex
correction: the shaded portion can include many gluon
lines, possibly
interacting with each other.

Fig 11. (a) The gluon in the vertex correction spans the
whole diagram. (b)
The gluon in the vertex correction spans only part of the
diagram. We see
that the diagram in (b) has one less loop than the one in (a).

Fig 12. The ``Ward identity" relating the vertex correction to
the self-energy.

Fig 13. A diagram contributing to the ``irrelevant"
factorized terms in
$N^2 G_c(z,w)$. Note the inner rim and the outer rim are
``decoupled" from
each other.


\begin{thebibliography}{99}

\bibitem{bz1} E. Br\'ezin and A. Zee, \np 402(FS), 613,
1993.

\bibitem{bz2} E. Br\'ezin and A. Zee, {\sl Compt.\ Rend.\
Acad.\
Sci.\/}
(Paris) t.317 II, 735 (1993).

\bibitem{bz3} E. Br\'ezin and A. Zee, Phys. Rev. {\bf E49}
(1994) 2588.



\bibitem{bz4} E. Br\'ezin and A. Zee, \np B424(FS), 435,
1994.

\bibitem{WIG} E. Wigner, {\sl Can.\ Math.\ Congr.\ Proc.\/}
p.174
(University of
Toronto Press) and other papers reprinted in Porter, op. cit.

\bibitem{POR} C.E. Porter, {\it Statistical\ Theories\ of
 \
Spectra:\ \
Fluctuations\/}
(Academic Press, New York, 1965).



\bibitem{MEH} M.L. Mehta, {\it Random\ Matrices\/}
(Academic
Press, New
York,
1991).

\bibitem{zee} A. Zee, ``Determinism plus chance in random
matrix theory",
SBITP preprint 95-43.


\bibitem{bhz} E. Br\'ezin, S. Hikami and A. Zee,
           Paris-Tokyo-Santa Barbara preprint (1994) LPENS-
94-35,
NSF-ITP-94-135, UT-
KOMABA-94-21. (hep-th.9412230).


\bibitem{danna} J. D'Anna, E. Br\'ezin, and A. Zee, {\sl Nucl.
Phys.} FS, Nucl. Phys. B 443 (1995) 433.

\bibitem{gross} R. Gopakumar and D. J. Gross,  Princeton
preprint PUPT-1520,
1994.

\bibitem{brody} T. A. Brody, J. Flores, J. B.  French, P. A.
Mello, A.
Pandey, and S. S. M. Wong, Rev. Mod. Phys. 53, 385, 1981.

\bibitem{verb} J. Verbaarschot, H. A. Weidenm\"uller, and M.
Zirnbauer,
Ann. of Phys. 153, 367, 1984.

\bibitem{douglas} M. Douglas, Rutgers preprint, hep-
th/9409098, 1994.

\bibitem{lattice} E. Br\'ezin and A. Zee, Nucl. Phys. (FS)
B441, 409, (1995).

\bibitem{li} M. Douglas and M. Li, Rutgers preprint, 1995.

\bibitem{voi} D. V. Voiculescu, K. J. Dykema, and A. Nica, {\it
Free\ Random\
Variables\/} (AMS, Providence, R. I.,
1992).

\bibitem{hag} U. Haagerup, private communication and to
be published.

\bibitem{waterloo} Lectures at the workshop on ``Operator
Algebra Free
Products and Random Matrices," Fields Insitute, March
1995, to appear in
the Proceedings.


\bibitem{doug} It turns out that there is a simple
interpretation of this
rule in
the language of string theory. Our wheel diagram can be
thought
as the propagation of a closed string, from the inner rim to
the outer rim
say. The annulus is then the worldsheet swept out by the
propagation. The
rule is that gluons must lie on the worldsheet, that is, the
annulus. We
see that the gluons in the vertex corrections can stay inside
the annulus
and not cross any gluon lines only by ``spanning" the whole
amplitude. We
thank M. Douglas for this remark. The reader not familiar
with this
language can safely ignore this footnote.

\bibitem{koba} T. S. Kobayakawa, Y. Hatsugai, M. Kohmoto,
and A. Zee,
ISSP-Tokyo preprint (1994)

\bibitem{gamma} It is tedious to calculate $\Gamma(z,w)$
perturbatively.
For the record, for $V(\varphi)=m^2 \varphi^2/2 +
g\varphi^4/4$,  the
undressed $\Gamma$ is given to $O(g^2)$ by
\bqy
\Gamma(z,w)&=& 1 -g(2+{1\over z^2}+{1\over zw}+{1\over
w^2})+g^2
(9+10({1\over
z^2}+{1\over zw}+{1\over w^2})\nonumber\\
&&+3({1\over z^4}+{1\over z^3 w}+{1\over z^2
w^2}+{1\over z w^3}
+{1\over w^4})
\eqy
It is a theorem that the coefficient of $1\over z^m w^n$
depends only on $m+n$.


\bibitem{thoo} G. 't Hooft, \np B72, 461, 1974.

\bibitem{bee} C.W.J. Beenakker,  \np B422(FS), 515, 1994.


\bibitem{eyn}   B. Eynard,  {\sl Nucl. Phys.} FS, in print,
1994.

\bibitem{for} P. J. Forrester, \np B435(FS), 421, 1995.

\bibitem{jer} See for instance, {\it Two\ Dimensional\
Quantum\ Gravity\
and\ Random\ Surfaces\/}, edited by D.J.
Gross and T.
Piran (World Scientific, Singapore, 1992).


\bibitem{BIPZ} E. Br\'ezin, C. Itzykson, G. Parisi, and J.B.
Zuber,
\journal  Comm. Math. Phys., 59, 35, 1978.

\bibitem{PAS} L.A. Pastur, \journal Theo. Math. Phys., 10,
67,
1972.

\bibitem{hz} S. Hikami and A. Zee, Nucl. Phys. B 446 (1995)
337.

\bibitem{nuclearguys} T. A. Brody, J. Flores, J. B. French, P. A. Mello, A.
Pandey, and S. S. M. Wong, Rev. Mod. Phys. 53, 385 (1981).

\bibitem{germans} J. Verbaarschot, H. A. Weidenmuller, and M. Zirnbauer,
Ann. of Physics 153, 367 (1984).

\end{thebibliography}
\end{document}